# Scanning Electron Microscopy and Metabolite Measurement Revealed the Stress Mechanism of PS-COOH Microplastics on *Rhodotorula mucilaginosa* AN5

Jiahao Ma (0000-0002-8978-3405, 2190790308@stu.hit.edu.cn), Xiangfei Meng, Zixin Li, Lexian Li, Jiwen Xu & Guangfeng Kan[*] (0000-0002-5357-7204, gfkan@hit.edu.cn)

School of Marine Science and Technology, Harbin Institute of Technology, Weihai, China

**Abstract**: Microplastics in the marine environment have been paid more and more attention by researchers, and the impact of these substances on marine microorganisms can not be ignored. Studies have shown that PS-COOH Microplastics are harmful to marine molluscs, algae and monads. This study explore the effect and mechanism of microplastics (80 nm PS-COOH) on Antarctic marine yeast, *Rhodotorula mucilaginosa* AN5 by bacterial count, Scanning Electron Microscopy (SEM) and metabolite analysis. The results illustrates that a 50 mg/L concentration of PS-COOH could inhibit 36.15±4.58% growth of yeast cells and 10 mg/L inhibit 80.20±3.27%. Microplastics stress causes changes in the content of some oxidative stress substances, including reactive oxygen species (ROS) 42.86% , malondialdehyde (MDA) 54.06% content and the activities of antioxidant enzymes such as catalase (CAT) 36.00% , peroxidase (POD) 66.67% and superoxide dismutase (SOD) 25.40%. These results revealed the possible stress effect of microplastic pollution on marine yeast and may affect bottom layer of marine ecosystem.

**Acknowledgments**

This work has been supported by The Science and Technology Project Fund for Undergraduate Students (Batch: 2021-2022 and 2020-2021) of Harbin University of Technology (Weihai), Weihai, China. At the same time, it is also supported by the experimental facilities and platform of the school of marine science and technology of Harbin University of Technology (Weihai), Weihai, China. Thanks Touch Education Technology Inc. for the help on the language improvement.

## 1. Introduction

Microplastics widely exist in the marine environment. They interact with other marine persistent pollutants and pollute marine biota when ingested (do Sul et al., 2014), which can be enlarged by the food chain and food web (Miloloža, M. et al., 2022). Therefore, the impact of microplastics on the bottom organisms of the marine food chain is considered to be an important topic (M Carbery et al., 2018). The earlier research on this topic originated from Fendall's research on microplastics in facial cleansers, which found that microplastics can be ingested by plankton and endanger the whole food chain and even mankind itself (Fendall et al., 2009). Subsequently, the interaction between a series of marine microorganisms including adhesion and plastic pollutants was also found(Gregory et al., 2009). Recently, more and more studies have been made on the relationship between microplastics and microorganisms in the ocean, including attaching (Ukil et al., 2022), fragmentation (Karkanorachaki et al., 2022), displacing (Song et al., 2022), and so on. Moreover, a study on microplastic stress *Brachionus Korea* and its oxidative stress was widely cited after it was published in 2016. It is noteworthy that there are few studies on the original growth state, cell morphology and membrane structure of microorganisms, especially yeast, a typical single-celled organism.

Marine *Rhodotorula* yeasts are dominant species in the ocean, which are widely distributed in the sea area including Antarctica and are closely related to the marine ecological environment (Kutty et al., 2008). This kind of yeast is also used in the food and chemical industry, which can produce many raw materials and products, including single-cell oil (Li et al., 2010) and acid protease (Lario et al., 2015). The Antarctic environment is considered to have extreme environmental characteristics such as low temperature, high salt, low light, oligotrophic, strong radiation and so on, which breed rich low-temperature microbial resources, including a variety of red yeast, which can produce unique metabolites. Therefore, the study of microorganisms from Antarctica is conducive to making better use of extreme environmental microbial resources.

Carboxylic Polystyrene (PS-COOH), one of the most common microplastic in marine (Grassi et al., 2020), is used in batteries, wearable devices and special materials. Its precursor, polystyrene (PS) and particularly fragmented PS foam are among the most common polymers observed during beach surveys in both sub-Antarctic and Antarctic regions (reviewed in Ivar do Sul et al. 2011) and recently found in sediments of the Ross Sea (Munari et al. 2017). There are few reports about the stress of this substance on microorganisms.PS-COOH has been used by researchers to study the interaction between organisms and microplastics (Zhang et al., 2021). Due to its special biological affinity and toxicity, PS-COOH is considered to be possible to stress marine microorganisms. Early studies suggested that these substances had significant biological toxicity to sea urchin embryos (Della et al., 2014). Although, above sevial researches provides the basis for micro plastic stress on molluscs, plants and monads, but there is little research on yeast.

In this study, we focused on the effect of PS-COOH on *Rhodotorula mucilaginosa* AN5, whose culture conditions, basic metabolites and stress effects have been proved to a certain extent in the past research of our laboratory (Kan et al., 2019, Zhang et al., 2022). After introducing the background of red yeast and marine microplastic pollution, the growth curve and cell morphology of *Rhodotorula mucilaginosa* AN5 under PS-COOH microplastic ball stress will be displayed, and the determinations of some oxidative stress substances are also be given. After that, the comparison between this work and other similar studies will be elaborate. Finally, the evaluation of the above work and suggestions for further research will be shown.

## 2. Materials and Methods

### 2.1 Microorganisms, Culture and Growth Curve

The yeast strain AN5 was isolated from Antarctic sea ice collected by the 23rd China Antarctic scientific expedition. Melting sea ice was progressively diluted and spread on 2216E agar plate at 10°C. After 7 days of culture, many bacteria and yeasts were obtained by single colony isolation. As a stress substance, 80 nm carboxyl polystyrene (PS-COOH) microspheres were from Tianjin Beiles chromatographic technology development center and stored in a refrigerator at 4 °C. Different amounts of PS-COOH were added to the YPDE liquid medium to make the final concentration of PS-COOH microspheres 10 mg / L and 50 mg / L respectively as the stress, which is the concentrations commonly adopted in previous studies (Nerland et al., 2016, He et al., 2022). The medium without PS-COOH microspheres was used as the control. The yeast AN5 was inoculated at 1% of the inoculation amount and cultured at 15 °C and 120 rpm for 8 days. Three replicates were set for each treatment. The number of yeast cells was measured by the blood cell counting method with the blank medium as the control.

## 2.2 Morphological Characteristics and Membrane Structure

Add PS-COOH with a final concentration of 10 mg / L to the culture medium, culture AN5 at 15 °C, take a small amount of culture solution at 3 days (logarithmic phase) and 7 days (stable phase) respectively, centrifuge at 12000 rpm for 3 min, precipitate the yeast, wash off the residual culture medium with a phosphate buffer solution at pH 7.8, and then fix, wash and gradient dehydration for standby.

The preliminary morphological analysis included the colony morphology and cell morphology referred to the past research (Kan et al., 2019). The yeast cells were fixed on small glass pieces and dehydrated with 10, 20, 30, 50, 70, and 90% ethanol solutions sequentially. The cell morphology was observed with a scanning electron microscope (SEM) (Hitachi, S-3400N Japan), and the acceleration voltage was constant at 5 kV.

## 2.3 Assay of oxidative stress indices

As 1 ml of logarithmic seed solution was inoculated in 100 ml YPD liquid medium, induced by 10 mg / L PS-COOH, and oscillatory cultured at 15 °C and 120 rpm for 8 days. Yeast without PS-COOH was used as the control group. Centrifugation at 8000 rpm for 10min, collect the cultured yeast, wash and weigh them, grind them with liquid nitrogen, melt them into 50 mM phosphate buffer solution (pH 7.8) overnight, then centrifugation at 12000 rpm for 30min, collect the supernatant and store it in the refrigerator at - 80 °C for the determination of the following steps.

Reactive oxygen species assay DCFH-DA (sodium 2 ', 7' - dichlorofluorescein diacetate; 2 ', 7' - dichlorofluorescein diacetate) at 5 mm was used as the mother liquor to achieve a working concentration of 10 μ m at 37 ° C and 50 rpm protected from light loading for 30 min. The fluorescence values were determined with a fluorescence microplate reader at excitation wavelength 485 nm and emission wavelength 528 nm. The ROS content was expressed by the fluorescence value of the extract solution corresponding to each mg of bacteriophage.

Malondialdehyde (MDA) content, peroxidase activity (POD) and catalase (CAT) activity were determined by the kit of Nanjing Jiancheng biological company; Superoxide (SOD) dismutase activity was determined by using the superoxide dismutase activity determination kit of Biyuntian biotechnology company, all above include three parallel tests.

2.4 Statistical analysis of data

Data were performed with at least three replicates and expressed as the mean value ± SD. Difference significance was analyzed by SPSS 17.0 (SPSS Inc., Chicago, USA). If $p < 0.05$, results were considered to be a statistically significant difference.

# 3. Results

3.1 growth curve

The resistance of AN5 to PS-COOH stress at different concentrations is shown in Figure 3-1. Under the stress of 10 mg / L and 50 mg / L PS-COOH, the growth of yeast was significantly inhibited and the growth rate was significantly slowed down, and the stress degree of the two concentrations on the growth of yeast was basically the same. The figure illustrates that a 50 mg/L concentration of PS-COOH could inhibit $36.15 \pm 4.58\%$ growth of yeast cells and 10 mg/L inhibit $80.20 \pm 3.27\%$. In addition, the secondary growth was more obvious in the presence of PS-COOH plastic microspheres.

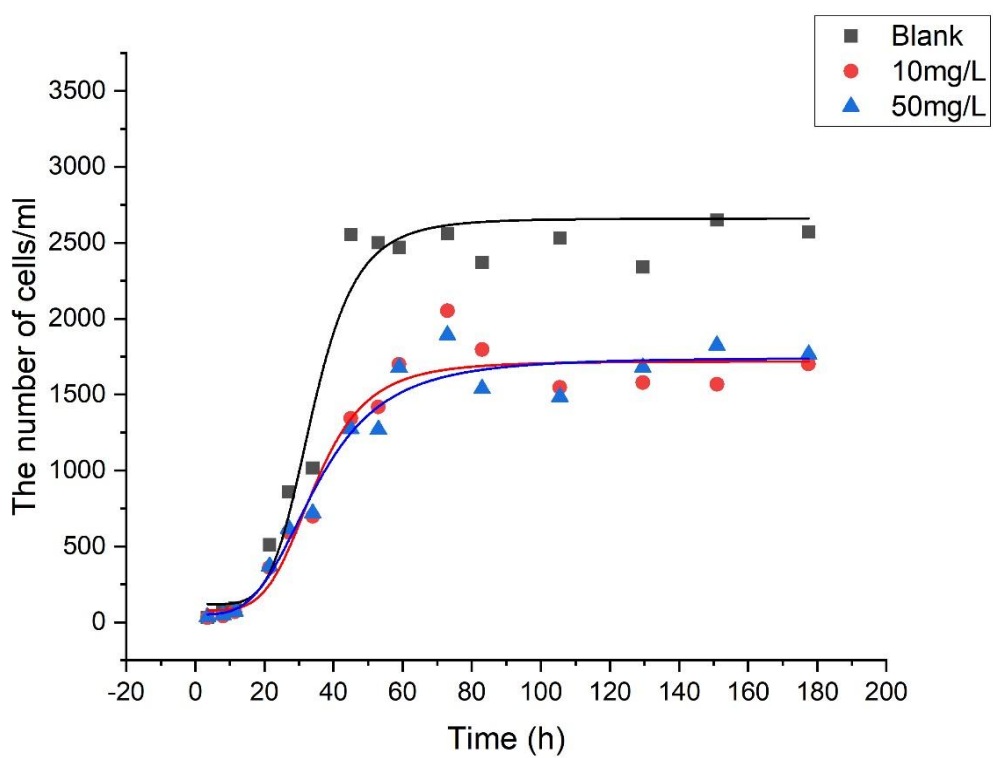

Fig. 3.1 Growth curve under stress

3.2 Morphology and membrane structure

The scanning electron microscope showed that when the yeast was in the logarithmic growth period, the surface of the control cell was smooth and full. And most of the cells were in the budding state. Under PS-COOH treatment, the surface of the yeast became rough and many depressions appeared, and obvious extracellular secretions could be seen. Many cells adhered together, and fewer cells were in the budding state than in the control group. When the yeast is in the stable growth period, the surface roughness of yeast cells increases, the adhesion between yeast cells increases. Additionally, PS-COOH Interact with the cell surface. So, fragments of cell death and rupture appear.

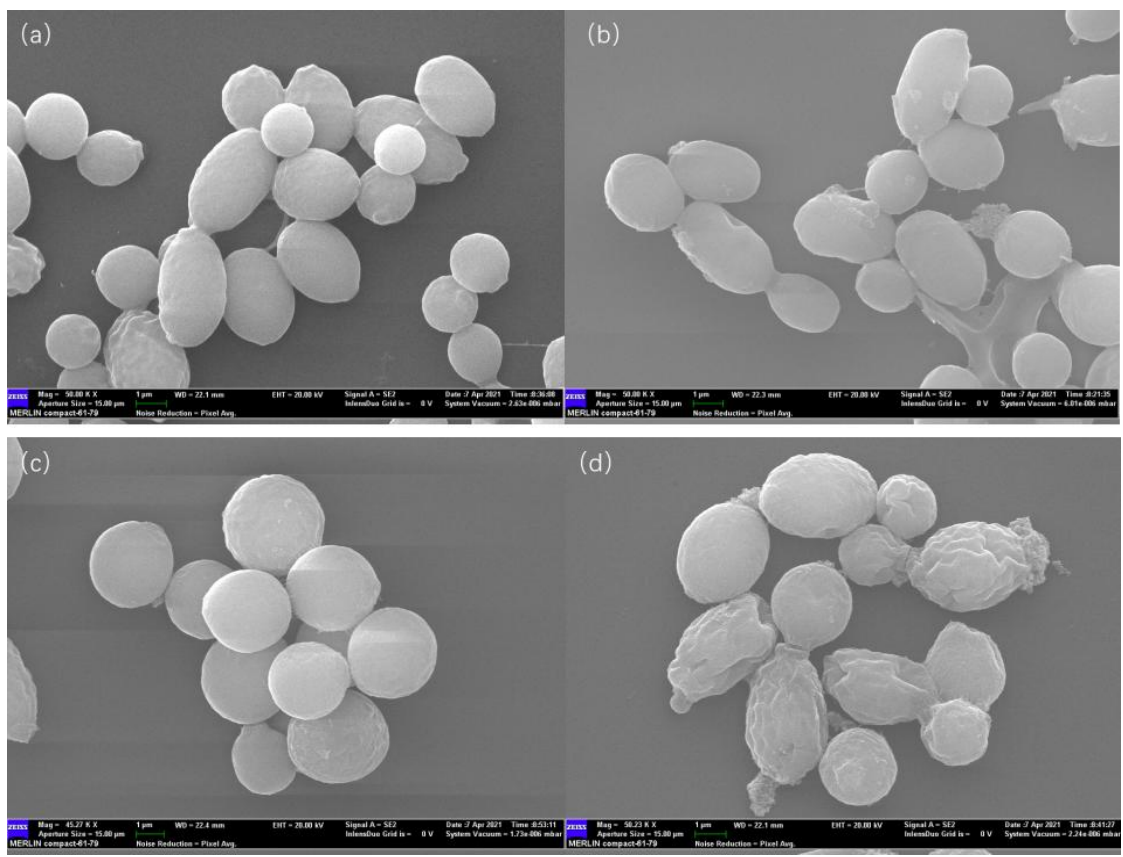

Fig. 3.2 morphological changes of *R. mucilaginosa* AN5 under PS-COOH stress
(a) The control cell grew for 3 days; (b) Under PS-COOH stress, the yeast grew for 3 days;
(c) The control cell grew for 7 days; (d) The strain grew for 7 days under PS-COOH stress.

3.3 Oxidative stress substances and enzyme activities

As shown in Figure 3.3, under stress, MDA content, CAT activity, POD activity and SOD activity increased in varying degrees, and CAT was the most significant. Microplastics stress causes changes in the content of some oxidative stress substances, including reactive oxygen species (ROS) 42.86%, malondialdehyde (MDA) 54.06% content and the activities of antioxidant enzymes such as catalase (CAT) 36.00%, peroxidase (POD) 66.67% and superoxide dismutase (SOD) 25.40%. Meanwhile, the content of ROS first increased and then decreased, which was lower than that of the control group after 6 days. These results reveal the oxidative stress produced by microplastics and the coping mechanism of yeast cells.

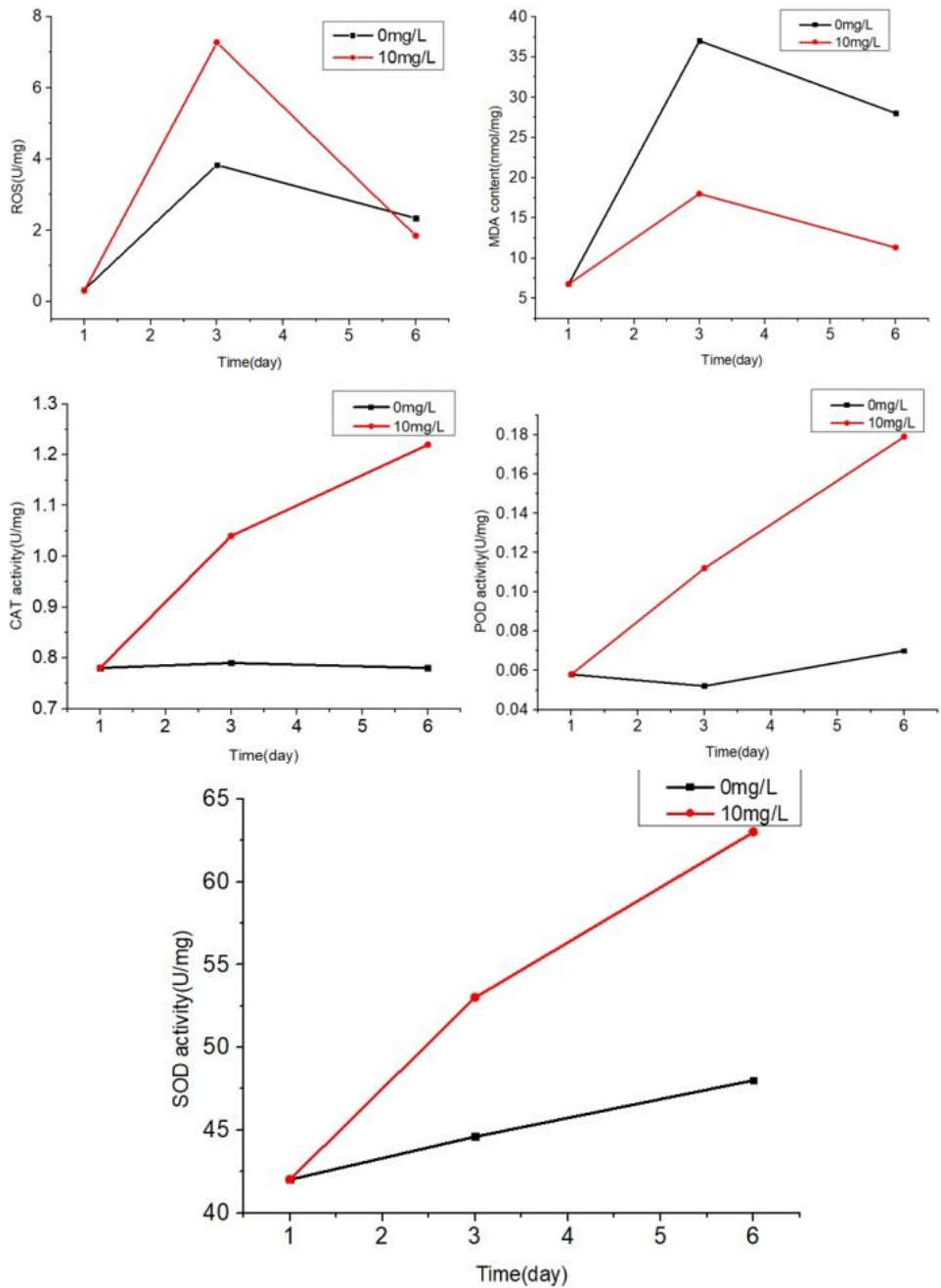

Fig.3.3 Oxidative stress substances and related enzyme activities

## 4. Discussion

Based on the growth curve, 80 nm carboxyl polystyrene microspheres inhibited the growth rate of R. mucilaginosa AN5 and reduced the maximum environmental capacity. The effect on the maximum environmental capacity did not change with the concentration, but the inhibition of the growth rate increased with the increase of concentration. According to the results of the scanning electron microscope, polystyrene plastic microspheres affected the normal state of yeast cell morphology and membrane structure. After that, the conjecture that reactive oxygen species damage the cell membrane structure was put forward, which was confirmed by the results of the measurement of reactive oxygen species and oxidative stress kinase activity.

Based on the above conclusions, the following conjecture can be put forward: with the increase of concentration, some microplastics will inhibit the growth of yeast widely existing in the ocean. This is because the microplastics with a small radius interact with the cell membrane structure, the cell membrane structure is damaged, and then the membrane lipid peroxidation results in the oxidative stress reaction of cells.

The inhibition of microplastics on the growth and reproduction of marine microorganisms has been reported by many fresh studies, including Wang's research (Wang et al., 2022) and Jachimowicz's research (Jachimowicz et al., 2022) in the field of biofilm and Sun's research on prokaryote (Sun et al., 2018). These studies are basically consistent with the inhibition phenomenon reported by us. From a comprehensive point of view, it can be concluded that "smaller particle size" plastics inhibit the growth of some microorganisms. In the existing studies, TEM more effectively observed the changes in extracellular substances of microbial cells (Sun et al., 2018). We used SEM to better observe the differences on the cell itself and membrane morphology. The results of reactive oxygen species and oxidative stress-related enzyme activities are similar to those on Halomonas alkaliphila, but the phenomenon of ROS decline in the stress group in the later stage is not found in the existing research(Zhang et al., 2017). This may be caused by secondary growth, or it may be caused by experimental operation or reagent kit.

Although there have been many studies on the microplastic stress of animals, plants and bacteria, the research on yeast is still relatively rare. This work revealed the effect of 80nm carboxyl polystyrene on Antarctica *Rhodotorula* and verified the conjecture about the mechanism of reactive oxygen species stress, which can provide a reference for the follow-up study of ecology and pollutants. However, the determination of physiological and biochemical indexes of yeast is less, and the research on the coping mechanism of yeast is not involved. For further research, proteome, transcriptome, metabolome, etc. it is necessary to reveal how yeast responds to microplastic stress through omics research. Other particle sizes and types of plastics can also be used in more research directions.

## 5. Conclusion

In the study of the relationship between microplastics and marine microorganisms, the stress relationship is less studied. Through three parts of growth curve mapping, scanning electron microscope observation and metabolite measurement, this study reached the conclusion that 80nm carboxyl polystyrene plastic microspheres stressed the population growth, cell morphology and oxidation emergency level of *R. mucilaginosa* AN5. These lay a foundation for further study on the impact of microplastic pollution on the marine ecological environment and biological population.

## Author information

### Affiliations

**Department of Marine Science and Technology, Harbin institute of Technology University, Shandong Province, 264209, China**

Jiahao Ma, Xiangfei Meng, Zixin Li, Lexian Li, Jiwen Xu & Guangfeng Kan*

### Contributions

LL and GK conceived and designed the whole study. JM, XM and JX were isolated and cultured and the growth curve was measured. LL and JM were observed by electron microscope. XM and ZL carried out some tests of metabolite measurement. JM wrote the manuscript, and GK and XM reviewed and proofread the manuscript. Everyone read the manuscript and contributed equally to the whole study.

# Ethics declarations

Conflict of interest

The authors report no declarations of interest.

Ethical statement

This article does not contain any studies with human participants or animal experiments by any of the authors.